# Exploring Generative AI assisted feedback writing for students' written responses to a physics conceptual question with prompt engineering and few-shot learning

Tong Wan[*] and Zhongzhou Chen[†]

Department of Physics, University of Central Florida, Orlando, Florida 32816, USA

**Abstract**

Instructor's feedback plays a critical role in students' development of conceptual understanding and reasoning skills. However, grading student written responses and providing personalized feedback can take a substantial amount of time, especially in large enrollment courses. In this study, we explore using GPT-3.5 to write feedback to student written responses to conceptual questions with prompt engineering and few-shot learning techniques. In stage one, we used a small portion (n=20) of the student responses on one conceptual question to iteratively train GPT to generate feedback. Four of the responses paired with human-written feedback were included in the prompt as examples for GPT. We tasked GPT to generate feedback to the other 16 responses, and we refined the prompt after several iterations. In stage two, we gave four student researchers (one graduate and three undergraduate) the 16 responses as well as two versions of feedback, one written by the authors and the other by GPT. Students were asked to rate the correctness and usefulness of each feedback, and to indicate which one was generated by GPT. The results showed that students tended to rate the feedback by human and GPT equally on correctness, but they all rated the feedback by GPT as more useful. Additionally, the successful rates of identifying GPT's feedback were low, ranging from 0.1 to 0.6. In stage three, we tasked GPT to generate feedback to the rest of the student responses (n=65). The feedback was rated by four instructors based on the extent of modification needed if they were to give the feedback to students. All the instructors rated approximately 70% (range from 68% to 78%) of the feedback statements needing only minor or no modification. This study demonstrated the feasibility of using Generative AI as an assistant to generate feedback for student written responses with only a relatively small number of examples in the prompt. An AI assistance can be one of the solutions to substantially reduce time spent on grading student written responses.

## I. Introduction

Developing conceptual understanding is one of the key learning goals in many introductory physics courses. There has been extensive research in investigating students' conceptual understanding and reasoning in introductory physics courses [1]. This body of research is often aligned with constructivist learning theory, which assumes that prior knowledge sets the foundation for new knowledge. Thus, effective teaching often requires eliciting and building on students' existing ideas [2].

Free-response format of conceptual questions in homework assignments provide students great opportunities to practice articulating reasoning and justifying conclusions. In addition, free-response format allows instructors to gain deep insights into student reasoning so that instructors can provide useful feedback and refine instruction when needed.

Instructor feedback plays a critical role in student learning. According to Ericsson's framework of "deliberate practice" [3], frequent feedback from an expert in addition to repeated and targeted practice are essential in acquiring expert performance. Moreover, empirical studies have confirmed that frequent feedback leads to substantial learning gains [4,5].

---

[*] tong.wan@ucf.edu
[†] zhongzhou.chen@ucf.edu

Although free-response questions provide opportunities for deliberate practice, grading and providing personalized feedback to students can be extremely time-consuming, especially in large enrollment courses. Institutions may not have enough resources to hire enough teaching assistants to grade homework assignments in large enrollment courses. Even when there is sufficient number of graders, the quality of grading and feedback can be inconsistent among individual graders without proper training. In response to these challenges and constraints, instructors may choose to give student credit for completion rather than correctness. However, without personalized feedback, students may not realize their mistakes even with instructor's solutions provided [6], let alone how to improve.

One possible way to overcome the challenge of grading and providing feedback to free-response questions is to off-load some of the grading tasks to generative artificial intelligence (AI), in particular large language models (LLMs). Recently, there are increased innovations of using generative AI in education [7]. Generative AI, especially LLMs, has been applied in a wide range of areas, such as personalized learning, intelligent tutoring, content creation, and essay grading (for a recent comprehensive review, see [8]).

There have been many recent attempts on using LLMs to grade or provide feedback to short answers or essays. Earlier efforts often utilize the pre-trained language models such as Bidirectional Encoder Representations from Transformers (BERT) [9–12]. However, those earlier language models require a significant amount of fine-tuning for satisfactory performance, a process that demand significant amount of technological expertise, computational resources, and large amounts of existing labeled text data (i.e., text data that are already classified), which creates obstacles for them to be widely adopted in teaching. With the rapid development in capability and enhanced accessibility of LLMs, more recent studies have focused on using more powerful LLMs such as GPT-3 to facilitate educational tasks, which are capable of achieving high performance with only simple prompt engineering (i.e., the process of developing and refining the prompt for LLMs to get desired output [13]).

We are only aware of one study that uses GPT to provide feedback on physics tasks. Steinert and colleagues [14] examined ChatGPT's capacity to provide formative feedback to students on a task that asks students to explain the underlying physics principle to an experiment. The study demonstrated that ChatGPT can be prompted to provide feedback that focuses on different theoretical aspects of learning, such as cognition, metacognition, and motivation. However, this study only included a single example student response, and it did not evaluate the quality of feedback.

In this paper we report our initial attempt at exploring the feasibility of using GPT-3.5 as a grading assistant for grading and providing feedback to students' written response to a single physics conceptual question. By combining our existing knowledge of students' preconceptions with simple prompt-engineering technique, we test if it would be possible to turn GPT-3.5 into an adequate grading assistant with only a few labeled data on student response, which could potentially save a significant amount of instructor grading effort, and provide useful feedback to students.

## II. Prior research on using LLM in education

LLMs such as GPT are being created to process and generate natural language. They are complex neural networks with billions of parameters that are pre-trained using a large corpus of text . A human user interacts with the LLM by inputting a piece of text, which is frequently referred to as a "prompt" in AI literature, The text can take any form. The LLM generates text output by predicting the most likely words to follow the prompt.

Multiple prior research have tested LLM's capacity to answer disciplinary questions. Kortemeyer [15], for example, tasked ChatGPT to complete all the course assignments like a human student would in an introductory physics course. With all the assignments graded and counted toward a final course grade, Kortemeyer found that ChatGPT can barely pass the course. Dahlkemper, Lahme, and Klein [16] evaluated students' perceptions of linguistic quality and scientific accuracy of ChatGPT responses to physics comprehension questions. Students' ratings for linguistic quality were essentially the same between ChatGPT and human-written solutions, but the ratings for scientific accuracy were much higher for human-written solution than ChatGPT responses. Moreover, the perceived difference in scientific accuracy decreased as the difficulty level of the questions increased.

Other studies have examined opportunities and limitations of LLM to generate disciplinary assessment questions. Kuchemann et al. [17] compared the quality of physics questions developed by pre-service teachers who used ChatGPT and those who used textbook. They found that the correctness and specificity (i.e., all relevant information to answer a question is present) of the questions were about the same between the two groups, but the clarity was much better in the questions developed by students who used a textbook. Additionally, pre-service teachers who used ChatGPT reported that ChatGPT is easy to use, but they felt neutral about usefulness and quality of ChatGPT's output.

### A. Prompt engineering

The quality of the output from an LLM may vary significantly depending on the quality of the prompt. Prompt engineering is the process of developing and refining a prompt in order to get desired output [13]. For example, Polverini and Gregoricic have shown that the performance of GPT on conceptual physics tasks can be significantly improved by using prompt engineering techniques [18].

As detailed in Polverini and Gregoricic, designing an effective prompt requires some understanding of how LLMs work, including its strengths and weaknesses. Since GPT is sensitive and responsive to context, the performance can be improved by providing the relevant context for response in the prompt. Providing context often include specifying the domain (e.g., specific topics or concepts) and specifying how to act (e.g., act like a physics teacher or an undergraduate student). However, Polverini and Gregoricic also cautioned that LLMs' context sensitivity and responsiveness can also be seen as a weakness. Shi et al. demonstrated that an LLM can be distracted by excess details and decreases its performance [19].

In addition, it has been widely documented that LLMs can sometimes generate outputs that are factually incorrect or contextually implausible (which is termed "hallucination" in AI research) [20]. Hallucination of LLMs can be effectively reduced by either employing better prompting engineering techniques or through domain specific fine-tuning of the model (see below). Yet because the output of LLMs is probabilistic, hallucinations of LLMs cannot be completely irradicated, which makes it critical for establishing human-in-the-loop procedures to minimize its potential negative impact. [8]

Moreover, research in other disciplines also indicates prompt engineering is an essential skill for leaners. Woo, Guo, and Susanto (Woo et al., 2023) explored English as a foreign language (EFL) students' prompt engineering pathways to writing using ChatGPT. The results suggest that prompt engineering is an important emergent skill for EFL students to improve writing. Heston and Khun discussed the necessity, challenges, and concerns of using prompt engineering techniques in medical education [22].

### B. Few-shot learning/prompting

Few-shot learning [23] is a framework in machine learning. It allows the pre-trained model to learn the underlying pattern from a few examples so that the model can generalize over new scenarios. Few-shot learning is often achieved by including in the prompt a few examples to demonstrate how LLMs should respond on a similar task, which is also called few-shot prompting. A notable recent example of using few-shot prompting in science education context is Zong and Krishnamachari [24], who used few-shot prompting to train GPT to solve math problems. They found that GPT was able to solve the problems with high accuracy, and the accuracy improved with increased number of examples.

### C. Chain-of-thought prompting

When the task requires multiple intermediate steps, chain-of-thought prompting [25] can be used to improve the accuracy of LLMs' response. A chain-of-thought prompt instructs an LLM to first generate a chain of intermediate reasoning steps before it generates the final answer. Chain-of-thought prompting can be combined with few-shot prompting to further enhance the LLM's performance. When combined, the examples provided in the prompt should include a chain of necessary intermediate reasoning steps, such as intermediate steps towards solving a math word problems.

### D. Fine-tuning

Fine-tuning is the process in which the parameters of an existing pre-trained LLM model is updated based on a large corpus of domain specific data, in order to improve performance on tasks of a specific domain. [26] In general, fine-tuning can achieve significantly greater performance improvement in the given domain compared to prompt-engineering. Latif and Zhai [27] have found that fine-tuned GPT is effective at automatic scoring student written responses to science questions. Zong and Krishnamachari [24] also found that fine-tuned GPT model outperformed GPT with prompt engineering only on math problem generation tasks. However, fine-tuning can be expensive as it requires large amounts of domain-specific, labeled data.

## III. Research questions

In this study, we use few-shot prompting to task GPT to generate feedback to students' written responses to one physics conceptual question. We investigate both student researchers' and physics instructors' perceptions of the quality of GPT-generated feedback. For students, we are interested in how they compare AI generated feedback to human generated feedback, in terms of both their perceived correctness and perceived usefulness. We are also interested in whether students can notice if a feedback statement is generated by AI. For physics instructors, we are investigating their perceptions of how much modification to the feedback, if any, is needed before they deem the feedback ready to be given to their students. This would give us insights into the extent to which an AI-based grading assistant could potentially save instructors' time on writing feedback. Specifically, we address the following three research questions (RQs):

RQ1: How do student researchers rate the correctness and usefulness of the GPT-generated feedback to student written response to a conceptual question?

RQ2: Can student researchers distinguish between human written and AI written feedback?

RQ3: How do instructors rate the level of editing needed to make the GPT-generated feedback satisfactory?

## IV. Methods

### A. Access to GPT-3.5 Turbo in complete mode

We used GPT-3.5-turbo in "complete" mode through Azure OpenAI Studio. The complete mode is different from the more popular "chat" mode, which is optimized for multiple rounds of conversation with a human, such as in ChatGPT. In complete mode, GPT model functions by treating the input prompt as an unfinished piece of text or a document, and generates output by predicting the most probably words and sentences could follow the prompt text. The complete mode is well-suited for tasks that require a single, well-structured response following instructions or prior examples. In contrast, in chat mode, GPT treats the prompt as a transcript of a conversation with a human, and tries to predict the response to the human. In this mode, GPT is much more likely to generate text such as "Sure! Here is what you requested". The chat mode is more suitable for multiple rounds of conversation. Since our purpose is to task GPT to only generate feedback to individual student written responses rather than having a conversation with the grader, the complete mode was chosen for our study.

### B. Question selection and feedback design

The student responses to the conceptual question were collected from an introductory physics course taught by the first author. The course was taught in a studio mode, in which lecture, recitation (discussion), and lab were integrated. The class enrollment was 99. During class, students were tasked to complete a tutorial that was adapted from the University of Maryland Open Source Tutorials. The tutorial targeted Newton's $2^{nd}$ and $3^{rd}$ laws for a multi-object system. Students received credit for completion, but their answers were not graded for correctness. Students submitted their written work online. A total of 85 student responses were collected from the class. The first author (also the instructor of record) typed down students' written responses before they were used to train GPT.

Question C from the Maryland tutorial was chosen for the current study, which is shown in Figure 1 together with questions A and B from the same tutorial. The scenario concerned a student pushes two boxes one next to another with a force of 200 newtons. The first two questions had students consider whether the accelerations of the boxes are the same, and then calculate the acceleration of the boxes. Question C asked whether box B is in danger of breaking given that the stuff in the box will break if the box experiences a 200 newtons force. The question explicitly asked students not to do any calculations, but answer intuitively and explain their reasoning.

In the tutorial, Question C was designed to elicit students' preconception about forces. The rest of the tutorial guides students to answer question C using Newton's second and third laws without doing calculations. Specifically, it prompts students to draw a free-body diagram for each box, and then asks students to tentatively assume that the force exerted by box A on box B equals 200 newtons and see where the assumption leads. By working through the tutorial, students are expected to recognize this assumption will lead to an incorrect and inconsistent conclusion that the net force on box A (as well as its acceleration) is zero, and thus they should reject this assumption. At the end of the tutorial, students are asked to refine their intuition about the central question (question C).

We chose question C because it elicits students' ideas as to whether a force can be "transmitted". We expect students' ideas to be rich in their variety and thus the question is well-suited as a test case for a potential Generative AI-based grading assistant.

The feedback that we intend for GenAI to generate includes a judgement statement on whether the students' conclusion and the reasoning are correct. If either the conclusion or reasoning was incorrect or missing, then feedback will directly point out what was incorrect and give a hint of an alternative direction of thought for the students to consider, which is the acceleration of the boxes (see Appendix A).

It is worth pointing out that this type of feedback is not designed to be given to students who are in the middle of completing the tutorial. Rather, a more likely use case would be when a question similar to question C is given again on following exams or homework, which will serve as an assessment on whether students have developed intuitions that are aligned with Newtonian physics after having completed the tutorial. If not, then the feedback could prompt students to think about Newtonian physics which they should have developed during the second half of the tutorial. Therefore, we designed the feedback to directly hint at a Newtonian physics reasoning, rather than trying to attend to and build on students' existing ideas in their response. Another reason for this choice is that instruction that builds on students' existing ideas (for example see studies based on the resource framework [28]) usually requires multi-round dialogue between the instructor and the student. This type of multi-round conversation is significantly more complicated to develop using generative AI, and out of the scope of the current exploratory study.

A student pushes two boxes, one in front of the other, as shown in the diagram. Box A has mass 75 kg, while box B has mass 25 kg. Fortunately for the student, the boxes are mounted on tiny rollers and slide with negligible friction. The student exerts a 200 N horizontal force on box A.

A. Without doing any calculations, state whether the acceleration of block A is greater than, less than, or equal to that of block B. How do you know?
B. Using any method you want, find the acceleration of the blocks. (Hint: It's possible to do this quickly and the blocks move together.)
C. **Box B contains kitchen stuff, including some poorly packed glassware that might break if the force pushing on the side of the box approaches 200 newtons. Recall that the student pushes on box A with a force of 200 newtons. Is that force "transmitted" to box B? In other words, is the glassware in the box in danger of breaking? Don't do any calculations; answer intuitively, and explain your thinking.**

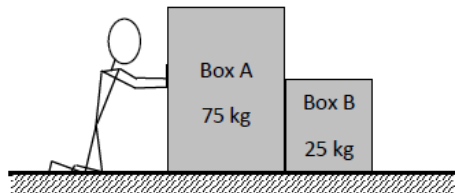

Figure 1. The first three questions in the tutorial. We used students' response to question C (in bold) in our study.

### C. Study Design

The study consisted of three stages. In Stage I, 4 student response and feedback pairs were used as examples to develop a prompt for the GPT model using few-shot prompting. The prompt was then refined to generate satisfactory feedback for 16 more student responses. In Stage II, four student raters were asked to rate both human generated and GPT generated feedback for the 16 responses. In stage III, the same prompt was used to generate feedback for the rest of the 65 responses, and four instructors were asked to rate the quality of the 65 feedback.

#### 1. Stage 1: GPT Assisted feedback generation with few-shot prompting

To prepare examples to include in few-shot learning, we first manually categorized all student responses. The responses were categorized based on whether the conclusion (as to whether or not the glassware in the box is in danger of breaking) is correct, and whether the explanation is correct, incomplete, incorrect or not present. We ended up with four categories: *correct conclusion with correct explanation*, *correct conclusion and incomplete/incorrect explanation, incorrect conclusion, and no explanation.* Since we did not see any students who gave correct reasoning but arrived at an incorrect conclusion, we did not have a category on that. Also, for those who did not give an explanation, we did not divide them based on whether the conclusion was correct. This is because we did not think two separate examples are necessary for those who did not give an explanation.

In each of the four categories, we selected five responses that we considered quite dissimilar to one another. The authors of the paper wrote feedback to those 20 responses, as if we were writing feedback to students in our own classes. We included one student response as well as the corresponding feedback from each category in the prompt given to GPT. We then tasked GPT to generate feedback to the rest of the 16 responses.

The prompt we developed include the following elements: the context (i.e., an instructor is giving feedback to student written responses), the physics problem, an expert response, physics concepts and principles involved, feedback instructions (i.e., what the feedback should look like), and four examples of student response and the feedback.

To generate a new feedback statement, the student response is appended to the end of the prompt, following the same structure as the four previous human generated examples. GPT then attempts to complete the text by predicting the most likely text that appears next, which is the feedback. Different from ChatGPT which is programmed to respond in a chat format, GPT in complete mode always generates only the feedback text (with the setting of a proper "stop sequence", which are specific character(s), such as a "new line" character, that signals GPT to stop generating following text).

The feedback generated by GPT in the first round had some obvious mistakes that resemble some of the common preconceptions identified in the PER literature. In an earlier study it was also documented that ChatGPT's outputs, which is the untrained, chat optimized version of GPT, reflects student preconceptions [29]. Therefore, we included in the prompt as contextual information those common preconceptions well documented in PER literature, such as "force can be transmitted through objects" and "force can be divided between objects" [30]. In addition to addressing preconceptions, we also made other changes, such as numbering the statements in the physics concepts and principles section and specifying what "not to do" in the feedback instructions. The prompt engineering process went through several iterations to optimize performance on the 16 selected responses. The final version of the prompt is shown in the Appendix.

All the 16 GPT-generated feedback were deemed to be correct by the authors [22]. By correct, we mean there was no apparent or indisputable mistakes, such as a misjudgment on the correctness of the student response, or a statement that resembles a student preconception. This is different from judging the feedback as potentially helpful to the students.

2. Stage 2: Student researcher evaluations

The 16 student responses and the corresponding human-written and GPT-generated feedback were given to four student raters who were involved in physics education research (PER). Three of the students were undergraduates and one was a first-year graduate student. All three undergraduate students had already completed the calculus-based Physics I course and excelled in the course. They were majoring in Aerospace Engineering, Computer Science, or Medical Laboratory Sciences. One of them was a learning assistant in Physics I during this study. All students were asked to evaluate the correctness and usefulness of each feedback, as well as indicated which one they think was generated by AI. The survey questions were shown in Figure 2.

1. Which of the feedback statements do you think is scientifically correct?
    a. A is correct.
    b. B is correct.
    c. Both are correct.
2. Which of the feedback statements do you think is more useful?
    a. A is more useful.
    b. B is more useful.
    c. Both are equally useful.
3. Which feedback do you think is generated by Generative AI (GPT)?
    a. A
    b. B
    c. Not sure

Figure 2. Survey questions given to student researchers to rate the 16 student responses with human-written and GPT-generated responses.

We note that the order of the responses was randomized rather than sequenced based on the categories. The order of feedback was also randomized such that GPT-generated feedback are not always listed first or second.

### 3. Stage 3: Instructor evaluations

In stage 3, we tasked GPT to generate feedback to the 65 responses that were not used in stage 1, based on the same prompt as we developed in stage 1. No iteration was performed to modify the prompt. The 65 feedback outputs were rated by four instructors, two of which were the authors of this paper, and the other two were non-PER faculty.

The goal of instructor ratings is to examine the potential of GPT to reduce the instructors' grading effort. Therefore, the scoring criteria is based on instructor's perceptions of how much modification is needed. All four instructors rated the feedback on a scale of 0 – 3. A score of 3 means that the instructor would give the feedback to a student without any modifications. This requires that the feedback is not only correct, but also addresses student's specific response. However, the feedback doesn't necessarily need to be address all the incorrect ideas in the response. A score of 2 means the feedback needs some quick modifications, such as some minor wording changes or a deletion of a part. A score of 1 means the feedback needs major revisions that often require deliberation and would take longer time to write than a feedback statement of score 2. Lastly, a score of 0 means the feedback needs to be rewritten completely. Those scoring conditions were fully communicated with the instructor raters.

## V. Results
### A. Student researcher ratings

We report student researchers' ratings on perceived scientific correctness and usefulness for both GPT-generated and human-written feedback to the 16 responses. We also show results for successful rates for identifying the GPT-generated feedback.

#### 1. Correctness

Figure 3 shows the distribution of correctness rating of GPT-generated and human-written feedback all four student researchers. Recall that all the feedback were considered correct (i.e., no apparent and indisputable mistakes) by the authors. Student A seemed to slightly favor feedback generated by GPT.

This student rated the both feedback as being correct for approximately half of the responses, and a greater number of GPT-generated feedback as being correct for the other half. In contrast, student B appeared to favor the feedback written by a human instructor. There was only one instance in which student B rated both feedback as being correct; the human-written feedback was rated by student B as correct much more often. Both students C and D did not show a clear preference. They rated both feedback as being correct for three quarters (or above) of the responses. Overall, we did not find a clear trend for favoring either human-written or GPT-generated feedback regarding correctness. It seemed both human-written and GPT-generated feedback was equally correct as perceived by the student researchers. The graduate student (student B) seemed to favor human generated feedback more.

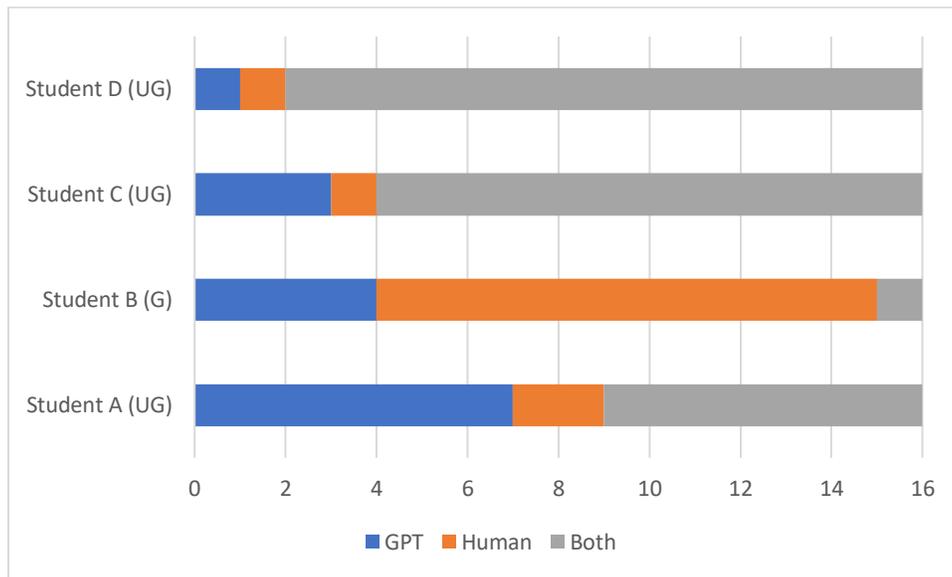

Figure 3. Distributions of correctness for GPT-generated and human-written feedback for all four student researchers.

2. Usefulness

Figure 4 shows the distribution of usefulness for GPT-generated and human-written feedback for all four student researchers. It appears all the student researchers rated the GPT-generated feedback as more useful. Student A showed the strongest preference toward GPT-generated feedback: three quarters of the GPT-generated feedback was rated more useful than human-written feedback. Even for student D, who had the highest frequency of both feedback statements being equally useful tended to favor GPT-generated over human-written feedback. Interestingly, student B who thought more human-written feedback statements were correct, still rated significantly more GPT-generated feedback statements as more useful.

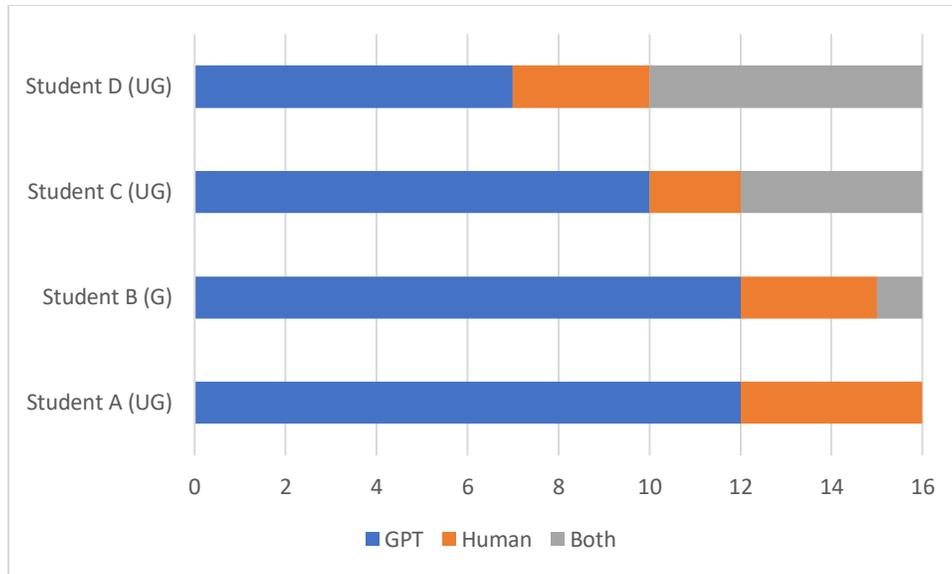

Figure 4. Distribution of usefulness for GPT-generated and human-written feedback for all four student researchers.

To gain insights into characteristics of the GPT-generated feedback that were perceived as more useful, we examined the feedback that were rated as more useful (or less useful) for all four student researchers. As shown in Table I, there were five instances in which the GPT-generated feedback was rated as more useful by all four student researchers, while only one instance in which the human-written feedback was unanimously rated as more useful.

At a quick glance, all five GPT-generated feedback that were perceived as more useful are much longer than the corresponding human-written feedback. Taking a closer look, GPT-generated feedback addresses students' responses to a greater extent. The five student responses happened to come from one of two categories: *correct conclusion with correct explanation*, and *no explanation.* GPT prompted students to take a step further in their reasoning when the explanation was correct; it gave a hint even when students did not provide an explanation.

For the GPT-generated feedback that was perceived as less useful, it appeared to address the student's response to the same extent as the human-written one did. Two of the student researchers rated that only the human-written feedback was correct, and the other two rated both human-written and GPT-generated feedback was correct. We infer that the lower rating in usefulness of the GPT-generated feedback was due to the lower perceived correctness.

Table I. List of GPT-generated feedback perceived as more useful or less useful by all four student researchers. The corresponding student responses and human-written feedback are also provided.

|  |  | Student response | Human-written feedback | GPT-generated feedback |
|---|---|---|---|---|
| GPT-generated feedback rated unanimously as more useful | | No because the force of box A on box b would not be 200 because box b has a smaller mass and would not take 200 N to move. | Both your conclusion and explanation are correct. | Your conclusion is correct and your explanation is on the right track. However, you may want to explain why smaller mass means smaller force. Think about an equation that relates force, mass and acceleration. |
| | | Yes, I think the glassware is in danger of breaking. | Your conclusion is incorrect. Please give an explanation so that we can provide some feedback on your reasoning. | Your conclusion is incorrect. You should think about why the force on box B may not be equal to the applied force of 200 N by the student on box A. Hint: Think about the mass and acceleration of the boxes. |
| | | Box B will be safe because it has a lower mass and requires less force to accelerate at the same rate as box A. | Both your conclusion and explanation are correct. | Your conclusion and explanation are both correct. To further strengthen your answer, think about the relationship between the force, mass, and acceleration in this scenario. |
| | | Yes | Your conclusion is incorrect. Please give an explanation so that we can provide some feedback on your reasoning. | Your conclusion is incorrect. Even though the student pushes on box A with 200 N, it does not mean that the box B receives a 200 N force. Think about the mass and acceleration of the two boxes and explain why the force on box B must be smaller than the 200 N applied force. |
| | | yes the glassware could become in danger of breaking. | Your conclusion is incorrect. Please give an explanation so that we can provide some feedback on your reasoning. | Your conclusion is incorrect. The force on box B is not necessarily equal to the force that the student applies to box A. Think about the mass and acceleration of these two boxes and consider what the force on box B must be to maintain the same acceleration. |
| Human-written feedback rated unanimously as more useful | | yes the force would be transmitted because they are moving together at 200 N. | Your conclusion is incorrect. The boxes moving together only means they have the same acceleration, not necessarily the same force. The 200 N force is only applied to box A, not box B. Will the force on box B be greater than or smaller than 200 N given its mass and acceleration? | Your conclusion is incorrect. It is true that the two boxes are moving together with the same acceleration, but the force on box B does not necessarily need to be 200 N. Consider how the masses of the two boxes impact the force acting on them. |

3. Which feedback is generated by AI?

Interestingly, although all student researchers tended to rate the GPT-generated feedback as more useful, the success rates of correctly identifying AI-generated feedback were much lower than chance, as shown in Table II. The graduate student researcher had a much higher success rate (63%) than the three undergraduate students had (ranged from 13% to 31%), but it is unclear to us why this was the case at this point.

It is worthwhile to point out that the current feedback is rather short, and follows a rather rigorous pattern, so it can be challenging for most people to distinguish between human and AI generated feedback. The purpose of this question is to test if the feedback created by AI seems natural and human-like to students at a first glance. If students could easily tell AI generated feedback from human generated feedback, it might cause them to have less trust in the former.

Table II. Distribution of GPT-generated feedback being indicated correctly and incorrectly.

|  | Correct | Incorrect | %Correct |
|---|---|---|---|
| Student A (UG) | 5 | 11 | 31% |
| Student B (G) | 10 | 6 | 63% |
| Student C (UG) | 2 | 14 | 13% |
| Student D (UG) | 3 | 13 | 19% |

### B. Instructor ratings

The distribution of instructor ratings is shown in Figure 5. Since the scoring is based on instructors' perceptions on how much modification the feedback is needed, we do not expect the instructors to reach an agreement on the individual feedback statements. Overall, the distribution across all four instructors appeared similar. Although the instructors were not quite aligned with the number of feedback statements that could be readily used with no modifications (score 3), they seemed to be consistent on the number of feedback statements that require either no modification (score 3), or only a minor modification (score 2). The percentages of feedback that requires a minor or no modification ranged from 68% to 78%. In addition, the average rating across all four instructors is 2.06 (on a scale of 0 - 3), which suggests that on average, the feedback only require minor modifications.

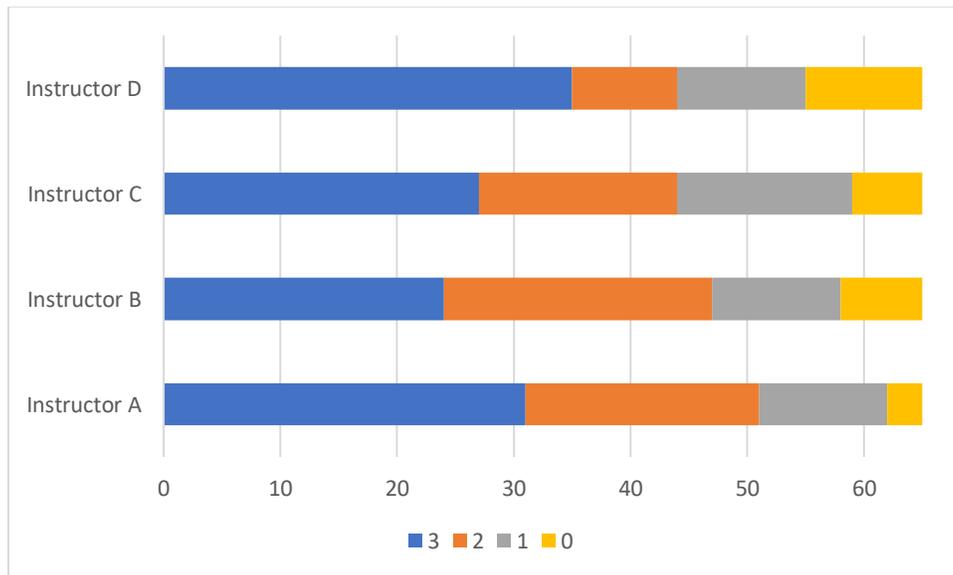

Figure 5. Distribution of ratings on a scale of 0 – 3 for all four instructors. The total number of GPT-generated feedback is 65.

There was one instance in which all four instructors agreed to use the feedback without any modification. The student stated, *"Yes, because both boxes will experience the same amount of force."* GPT's feedback stated, *"Your conclusion is incorrect. Both boxes may experience the same acceleration, but they need to have different forces depending on the mass. Think about this in terms of net force and the mass of box B relative to box A."* GPT correctly pointed out that the acceleration was the same, but the net forces are different because the boxes have difference mass.

The lowest rated GPT feedbacks received three "0" and one "2", which occurred in two instances. For example, a student stated, *"That force will have been transferred and the glassware will break."* GPT's feedback stated, *"Your conclusion is incorrect, but it seems like you have a good understanding of the concepts. Think more about why the force can't possibly be transferred to box B."* It is possible that GPT "interpreted" the students' expression of "force will have been transferred" as indicating that the force will actually not be transferred to box B.

## VI.     Discussion

In this study, we made an initial attempt to test the feasibility of using GPT to assist in grading and generating personalized feedback to student written responses to one conceptual question. Taking advantage of rich existing PER literature in student preconceptions on Newtonian mechanics, we included relevant common preconceptions in the prompt to improve the accuracy of feedback. We also categorized student responses and provided GPT one example response-feedback pair from each category to initiate the few-shot learning process.

The results showed that three out of four student researchers perceived GPT-generated and human-written feedback to be equally correct. At the same time, all the student researchers rated GPT-generated feedback as more useful. This was probably because GPT-generated feedback statements are generally longer and address students' responses to a greater extent, especially when students gave a correct conclusion with a correct explanation or when they did not provide an explanation. In contrast, the authors gave the same feedback to students whose responses were classified in either of the two categories. Our justification for the short feedback is that it saves time, and the saved time can be used to write longer feedback for students who the instructor judge to need more detailed feedback, such as those who gave incorrect conclusions or explanations. However, this practice could potentially leave students who gave the correct answer with the feeling of being neglected, which is where GPT can be extremely helpful as it treats every response with equal patience and effort.

More interestingly, the percentages of correctly identifying the GPT-generated feedback were low overall among the students, which suggests that the student researchers often perceive GPT generated feedback as human-like, and perceive brief human feedback as being created by a machine. A possible explanation might be that students expect the human expert to be more helpful, and a machine to be more repetitive. It is worth noting that the purpose here is not to evaluate student researchers' ability to distinguish between AI and human, as student researchers based their judgements on just a few lines of feedback. The outcomes might be different if the feedback were more extensive, or a multi-turn conversation is involved. For example, Jones and Bergen [31] showed that in a multi-turn conversation situation, GPT-3.5 is only misclassified as a human in 14% of the cases. However, the results do suggest that students are unlikely to quickly judge the AI generated feedback as "artificial" or "machine generated" at a first glance. In addition, our results also speak to one strength of generative AI, which is that it has infinite patience and will always respond to student answers with sufficient detail, unlike human graders whose performance can be easily impacted by fatigue and emotion, resulting in short and repetitive feedback that could seem more "machine like".

For the four instructors, only about 30% of the time would they consider major modification or re-writing of GPT generated feedback, indicating a relatively high percentage of satisfaction for GPT's performance in the current setting. However, it should be cautioned that the number of student answers is relatively small in this study, so it remains to be seen whether the 70% satisfactory rate can be generalized to more diverse student answers or when generating more sophisticated feedback. There are some studies that show that the performance of LLMs might be artificially boosted by the models relying on certain common keywords that exist in the specific dataset, or expected in the output (also known as short-cut learning [32])

Together, our results suggest that LLMs such as GPT-3.5 has promising potential for serving as a grading assistant with just prompt engineering and few-shot learning, without the need for additional fine-tuning. Even though the feedback is only satisfactory in 70% of the cases, it could still be a significant save in the amount of grading time and effort required from the instructors. This allows instructors to focus on addressing student answers that require more attention, and assign more open response problems to students. From students' point of view, GPT generated feedback statements are perceived as more useful, and sometimes even more "human-like" than actual human generated feedback. Moreover, GPT generated feedback has the unique advantage of being highly consistent in the level of detail regardless of the amount of grading workload, which could provide more equal feedback to all students.

## VII.    Limitations and future work

As an explorative initial study, there are many limitations and caveats that can be improved or investigated in future work.

First, in the current study, one of the authors manually categorized all the student answers and provided GPT with one solution/feedback pair from each category as example. The manual categorization process is very time consuming and impractical if LLMs are to be used as a grading assistant. Future studies could explore two possible alternate solutions. One of them is to randomly select a small number (for example 10 – 20) student answers and write feedback to those as initial examples. The other solution is to use LLM to perform clustering analysis on student responses using text embedding and machine learning (for example, see [33]). Example answer/feedback pairs can be created by selecting 1-2 student answers from in each response cluster.

Second, the current study only involves one conceptual question. Future studies could test similar approaches on a wider variety of problems and evaluate the performance. Moreover, a valuable future direction is to investigate whether GPT can be used to generate feedback for a class of similar problems involving the same physics principle, by providing a single, context independent general prompt. Future studies could also explore the possibility of using LLMs as chatbots, and engage students in multi-round conversation that builds on student's existing knowledge to construct new knowledge, based on theoretical frameworks such as the resource model [28].

Third, the feedback was only evaluated by four student researchers, which is a small number. Moreover, all of them are involved in PER projects, and two of them are actually involved with other projects involving LLM, which could have an influence on their judgements. In future studies, we plan to survey students who are enrolled in introductory physics courses about their perceptions of the AI-generated feedback they receive. We also plan to survey more graduate teaching assistances without prior knowledge of GenAI and evaluate their response to AI generated feedback.

Forth, we tasked instructors to rate the feedback based on their perception of the amount of editing needed before they deem the feedback ready for students. We did not provide more contextual information regarding the use of the problem and feedback, such as whether the problem was used as part of a homework assignment, exam, or in-class activity. Nor did we ask the instructors to rate the quality of the generated feedback. A content-specific rubric will need to be developed for instructors to rate the quality of the feedback.

Furthermore, the number of student responses in this study was still relatively small, which enabled the study to be conducted using GPT-3.5-turbo, which is a large 20B parameter model, and at minimum cost to the researcher. For future applications with much larger student body and many more problems, the usage cost of such application could be significant. Therefore, it is worth investigating whether much smaller models pre-trained on materials from a specific discipline or a specific course could reach a similar level of performance.

Finally, using Generative AI to assist grading has some additional potential caveats. For example, AI-generated feedback could be potentially biased [34,35]because LLMs like GPT are pre-trained from large amount of data from different sources including the internet and social media, and therefore its outputs can be influenced by existing bias contain in the source of its training. LLMs could also generate incoherent chain of argument or even factually incorrect information, often referred to as "hallucination" in AI literature [36–38]. Currently, we recommend instructors who intend to use GPT as a grading assistant check all the feedback and make edits if needed before sending the feedback to students. Future research should explore using AI or machine learning methods to automatically suggest the feedback that is most likely to be incorrect for instructors to review. Lastly, another issue worth considering regarding the use of AI to assistant grading is its environmental impact, as training and running AI systems often require substantial computing power and thus significant electricity consumption [39].

## VIII. Conclusion and implications

In conclusion, we believe that generative AI holds significant potential in serving as a grading assistant for open response questions. One possible model of using GPT as a grading assistant could be as follows: First, the instructor is required to write feedback on several student responses chosen by the assistant, and input any known student preconceptions from the literature. Second, the assistant grades and writes feedback on all student responses, using a prompt that incorporates the instructor input. Third, all response-feedback pairs are presented to the instructor, and the instructor edits the feedback which takes significantly less time than generating all feedback from scratch. Finally, the corrected feedback will be sent back to students, will be incorporated into the prompt for better future performance. Such a process could significantly reduce the grading load for instructors and increase feedback quality for students, potentially leading to improved student conceptual understanding.

**Acknowledgement**

The authors would like to thank Dr. Xiaofeng Feng and Dr. Richard Jerousek for assisting in rating the feedback. This work is supported in part by the University of Central Florida Digital Curriculum Innovation Initiative.

**Appendix**

Final version of GPT prompt

#Context
A physics instructor is rating students' answers to the following physics problem:
## Physics Problem:
A student pushes two boxes, one in front of the other, as shown in the diagram. Box A has mass 75 kg, while box B has mass 25 kg. Fortunately for the student, the boxes are mounted on tiny rollers and slide with negligible friction. The student exerts a 200 N horizontal force on box A.
Box B contains kitchen stuff, including some poorly packed glassware that might break if the force pushing on the side of the box approaches 200 newtons. Recall that the student pushes on box A with a force of 200 newtons. Is that force "transmitted" to box B? In other words, is the glassware in the box in danger of breaking? Don't do any calculations; answer intuitively, and explain your thinking.

# The instructor rates students' answer and gives feedback based on how similar it is to this expert answer:

## Expert Response:
No, the 200 N force is not 'transmitted" to box B. For box A, there is a 200 N applied force to the right and a force by box B to the left, thus the net force on box A is less than 200 N. Box A and box B have the same acceleration, but box B has a smaller mass. To maintain the same acceleration, net force on box B needs to be smaller than net force on box A. Therefore, the force on box B must be smaller than 200 N.

## Physics principles involved:
1. The net force on an object is equal to the mass times its acceleration.
2. "Force", "energy", and "power" are different concepts.
3. Force is never transmitted from one object to another.
4. Force is never divided among objects.
5. Force is never distributed among objects.
6. Force is never absorbed by an object.
7. Force cannot pass through an object.
8. Two forces acting on different bodies do not "cancel out".
9. Box B has smaller mass, so to maintain the same acceleration, force on box B needs to be smaller, which means that it must be smaller than 200 N.

# The feedback should start with whether the answer is correct or incorrect, followed by a justification, and then a follow up question for further thinking. The feedback should not give away the expert answer. Those physics principles should not all be repeated in the feedback to students.

# Here are several examples of student answer and instructor feedback:

Answer: No because in order to maintain the same acceleration, the force on box B must be lower than 200 N.

Feedback: Both your conclusion and explanation are correct.

Answer: No, the 200 N push force does not transfer to box B. The push force is only being applied to box A and you cannot assume that the force is transferred. The force acting on box B would be the force of A on B, not from the push force.

Feedback: Your conclusion is correct, but your explanation needs some improvements. You are correct that the force does not transfer to box B, but you will need to explain why force of A on B must be smaller than 200 N. Hint: Think about the mass and acceleration.

Answer: Yes, it will break because the forces are equal and opposite force on each object.

Feedback: Your conclusion is incorrect. Yes, the force by Box B on Box A is equal and opposite to the force by Box A on Box B. However, it does not necessarily mean that those forces are equal to 200 N applied force by the student on Box A. You may want to consider whether the force on box B is greater than or smaller than 200 N based on its mass and acceleration.

Answer: No, it is not in danger there isn't a 200 N force on the box.

Feedback: Your conclusion is correct, but you need to provide an explanation for why the force on box B is not 200 N.

Answer: [new student response goes in here.]

Feedback:

# References

[1] R. Beichner, *An Introduction to Physics Education Research*, in *Getting Started in PER*, edited by C. Henderson and K. Harper, Vol. 2 (2009).

[2] L. S. Shulman, *Those Who Understand: Knowledge Growth in Teaching*, Educational Researcher **15**, 4 (1986).

[3] K. A. Ericsson, R. T. Krampe, and C. Tesch-Romer, The Role of Deliberate Practice in the Acquisition of Expert Performance, 1993.

[4] Louis Deslaurier, Ellen Schelew, and Carl Wieman, *Improved Learning in a Large-Enrollment Class*, Science (1979) **332**, 858 (2011).

[5] P. Black and D. Wiliam, *Assessment and Classroom Learning*, Int J Phytoremediation **21**, 7 (1998).

[6] J. Larreamendy-Joerns, G. Leinhardt, and J. Corredor, *Six Online Statistics Courses: Examination and Review*, American Statistician.

[7] D. Baidoo-Anu and L. Owusu Ansah, Education in the Era of Generative Artificial Intelligence (AI): Understanding the Potential Benefits of ChatGPT in Promoting Teaching and Learning, 2023.

[8] E. Kasneci et al., *ChatGPT for Good? On Opportunities and Challenges of Large Language Models for Education*, Learn Individ Differ **103**, 102274 (2023).

[9] Z. Li, C. Zhang, Y. Jin, X. Cang, S. Puntambekar, and R. J. Passonneau, *Learning When to Defer to Humans for Short Answer Grading*, (n.d.).

[10] C. Sung, T. Ma, T. I. Dhamecha, V. Reddy, S. Saha, and R. Arora, *Pre-Training BERT on Domain Resources for Short Answer Grading*, EMNLP-IJCNLP 2019 - 2019 Conference on Empirical Methods in Natural Language Processing and 9th International Joint Conference on Natural Language Processing, Proceedings of the Conference 6071 (2019).

[11] A. Ahmed, A. Joorabchi, and M. Hayes, *On Deep Learning Approaches to Automated Assessment: Strategies for Short Answer Grading*, in *Proceedings of the 14th International Conference on Computer Supported Education* (SCITEPRESS - Science and Technology Publications, 2022), pp. 85–94.

[12] A. Condor, *Exploring Automatic Short Answer Grading as a Tool to Assist in Human Rating*, Lecture Notes in Computer Science (Including Subseries Lecture Notes in Artificial Intelligence and Lecture Notes in Bioinformatics) **12164 LNAI**, 74 (2020).

[13] P. Liu, W. Yuan, J. Fu, Z. Jiang, H. Hayashi, and G. Neubig, *Pre-Train, Prompt, and Predict: A Systematic Survey of Prompting Methods in Natural Language Processing*, (2021).

[14] S. Steinert, K. E. Avila, S. Ruzika, J. Kuhn, and S. Küchemann, *Harnessing Large Language Models to Enhance Self-Regulated Learning via Formative Feedback*, (2023).

[15] G. Kortemeyer, *Could an Artificial-Intelligence Agent Pass an Introductory Physics Course*, Phys Rev Phys Educ Res **19**, (2023).

[16] M. N. Dahlkemper, S. Z. Lahme, and P. Klein, *How Do Physics Students Evaluate Artificial Intelligence Responses on Comprehension Questions A Study on the Perceived Scientific Accuracy and Linguistic Quality of ChatGPT*, Phys Rev Phys Educ Res **19**, (2023).

[17] S. Küchemann, S. Steinert, N. Revenga, M. Schweinberger, Y. Dinc, K. E. Avila, and J. Kuhn, *Can ChatGPT Support Prospective Teachers in Physics Task Development?*, Phys Rev Phys Educ Res **19**, (2023).

[18] G. Polverini and B. Gregorcic, *How Understanding Large Language Models Can Inform the Use of ChatGPT in Physics Education*, Eur J Phys (2023).

[19] F. Shi, X. Chen, K. Misra, N. Scales, D. Dohan, E. Chi, N. Schärli, and D. Zhou, *Large Language Models Can Be Easily Distracted by Irrelevant Context*, (2023).

[20] M. Lee, *A Mathematical Investigation of Hallucination and Creativity in GPT Models*, Mathematics **11**, (2023).

[21] D. J. Woo, K. Guo, and H. Susanto, Cases of EFL Secondary Students' Prompt Engineering Pathways to Complete a Writing Task with ChatGPT, 2023.

[22] T. F. Heston and C. Khun, *Prompt Engineering in Medical Education*, International Medical Education **2**, 198 (2023).

[23] Y. Wang, Q. Yao, J. T. Kwok, and L. M. Ni, *Generalizing from a Few Examples: A Survey on Few-Shot Learning*, ACM Comput Surv **53**, (2020).

[24] M. Zong and B. Krishnamachari, Solving Math Word Problems Concerning Systems of Equations with GPT-3, 2023.

[25] J. Wei, X. Wang, D. Schuurmans, M. Bosma, B. Ichter, F. Xia, E. Chi, Q. Le, and D. Zhou, *Chain-of-Thought Prompting Elicits Reasoning in Large Language Models*, (2022).

[26] E. Radiya-Dixit and X. Wang, *How Fine Can Fine-Tuning Be? Learning Efficient Language Models*, in *Proceedings of the Twenty Third International Conference on Artificial Intelligence and Statistics*, edited by S. Chiappa and R. Calandra (PMLR, 2020), pp. 2435–2443.

[27] E. Latif and X. Zhai, *Fine-Tuning ChatGPT for Automatic Scoring*, (2023).

[28] D. Hammer, *Student Resources for Learning Introductory Physics*, Am J Phys **68**, S52 (2000).

[29] S. Wheeler and R. E. Scherr, *ChatGPT Reflects Student Misconceptions in Physics*, in *2023 Physics Education Research Conference Proceedings* (American Association of Physics Teachers, 2023), pp. 386–390.

[30] R. E. Scherr and E. F. Redish, *Newton's Zeroth Law: Learning from Listening to Our Students*, Phys Teach **43**, 41 (2005).

[31]   C. Jones and B. Bergen, *Does GPT-4 Pass the Turing Test?*, (2023).

[32]   M. Mitchell and D. C. Krakauer, *The Debate over Understanding in AI's Large Language Models*, Proceedings of the National Academy of Sciences **120**, e2215907120 (2023).

[33]   P. Tschisgale, P. Wulff, and M. Kubsch, *Integrating Artificial Intelligence-Based Methods into Qualitative Research in Physics Education Research: A Case for Computational Grounded Theory*, Phys Rev Phys Educ Res **19**, 020123 (2023).

[34]   D. Rozado, *The Political Biases of ChatGPT*, Social Sciences 2023, Vol. 12, Page 148 **12**, 148 (2023).

[35]   L. Lucy and D. Bamman, *Gender and Representation Bias in GPT-3 Generated Stories*, 48 (2021).

[36]   M. Zhang, O. Press, W. Merrill, A. Liu, N. A. Smith, and P. G. Allen, *How Language Model Hallucinations Can Snowball*, (2023).

[37]   D. Bilbao, A. Gelbukh, A. Rodrigo, and M. Lee, *A Mathematical Investigation of Hallucination and Creativity in GPT Models*, Mathematics 2023, Vol. 11, Page 2320 **11**, 2320 (2023).

[38]   Y. Bang et al., *A Multitask, Multilingual, Multimodal Evaluation of ChatGPT on Reasoning, Hallucination, and Interactivity*, (2023).

[39]   E. M. Bender, T. Gebru, A. McMillan-Major, and S. Shmitchell, *On the Dangers of Stochastic Parrots: Can Language Models Be Too Big?*, FAccT 2021 - Proceedings of the 2021 ACM Conference on Fairness, Accountability, and Transparency 610 (2021).